\documentstyle[12pt]{article}
\newcommand{\BF}[1]{\mbox{\boldmath $#1$}}

\def\abstracts#1#2#3{{
        \centering{\begin{minipage}{4.62in}\baselineskip=13pt
        \small
        \centerline{\bf Abstract}
        \vspace*{0.2cm}                
        \parindent=0pt #1\par
        \parindent=18pt #2\par
        \parindent=15pt #3
        \end{minipage} }\par}}
\renewcommand{\thefootnote}{\fnsymbol{footnote}}
\begin{document}
\vspace*{-2cm}
\hfill \parbox{4cm}{ FUB-HEP 19/92 \\
               HLRZ Preprint 77/92 \\
               ~ }\\
\vspace*{2cm}
\centerline{\LARGE \bf Critical Exponents of the Classical}\\[0.2cm]
\centerline{\LARGE \bf 3D Heisenberg Model:}\\[0.4cm]
\centerline{\LARGE \bf A Single-Cluster Monte Carlo
Study\footnotemark}\\[0.3cm]
\footnotetext{\noindent Work supported in part by Deutsche
Forschungsgemeinschaft under grant Kl256.}
\addtocounter{footnote}{-1}
\renewcommand{\thefootnote}{\arabic{footnote}}
\vspace*{0.2cm}
\centerline{\large {\em Christian Holm\/}$^1$ and
                   {\em Wolfhard Janke\/}$^2$}\\[0.4cm]
\centerline{\large    $^1$ {\small Institut f\"{u}r Theoretische Physik,
                      Freie Universit\"{a}t Berlin}}
\centerline{    {\small Arnimallee 14, D-1000 Berlin 33, Germany}}\\[0.2cm]
\centerline{\large    $^2$ {\small H\"ochstleistungsrechenzentrum,
                      Forschungszentrum J\"ulich}}
\centerline{    {\small Postfach 1913, D-5170 J\"ulich, Germany }}\\[0.5cm]
\abstracts{}{
We have simulated the three-dimensional Heisenberg model
on simple cubic lattices, using the single-cluster Monte Carlo update
algorithm.
The expected
pronounced reduction of critical slowing down at the phase transition is
verified. This
allows simulations on significantly larger lattices than in previous
studies and consequently a better control over systematic errors. In one set of
simulations we employ
the usual finite-size scaling methods to compute the critical exponents
$\nu,\alpha,\beta,\gamma,\eta$  from a few measurements in the vicinity
of the critical point,
making extensive use of
histogram reweighting and optimization
techniques.
In another set of simulations we report measurements of improved estimators
for the
spatial correlation length and the susceptibility in the high-temperature
phase, obtained on lattices with up to $100^3$ spins. This enables us to
compute
independent estimates of $\nu$
and $\gamma$ from power-law fits of their critical divergencies.
}{}
\thispagestyle{empty}
\newpage
\pagenumbering{arabic}
%
                     \section{Introduction}
%
Critical exponents are the distinguishing parameters characterizing
continuous phase transitions. Most theoretical estimates of their values
have been
calculated along three different routes. First, assuming universality,
from (resummed) perturbation expansions of generic field theoretical
models. Second, from high-temperature series expansions of lattice models,
and third, from Monte Carlo (MC) simulations of the associated Boltzmann
distributions. All these approaches are based on resummation or
extrapolation techniques whose systematic errors are difficult to
control. To gain confidence in the numerical values of the exponents
it is therefore quite important to have several independent approaches
for cross-checks available.

Until a few years ago the precision of Monte
Carlo estimates was comparatively poor, since this
approach was plagued not only by
systematic but also by statistical errors. The problem is that MC
algorithms based on {\em local\/} update procedures are severely
hampered near criticality by extremely long autocorrelation
times (so-called critical slowing down) which reduce the effective statistics
and thus increase the statistical errors considerably \cite{r2}.
The recent development \cite{r1} of {\em global\/} update algorithms that
overcome the problem of critical slowing down is a major step forward.
It is this algorithmic improvement combined with the higher speed of modern
computers that makes systematic tests of
finite-size scaling (FSS) predictions \cite{barber} and a reliable computation
of critical exponents much more feasible than a few years ago.

For spin systems cluster algorithms \cite{r3,r4} turned
out to be particularly successful. Recent applications of the multiple
\cite{r3} and single \cite{r4} cluster variants to two-dimensional (2D)
Ising  \cite{r5,tamayo,r6}, XY \cite{r8,r7,r9,r9a},
Heisenberg \cite{r10,r10a} and
other  $O(n)$ \cite{r11} models have demonstrated that both
variants are equally good tools to simulate these 2D systems. In three
dimensions (3D), however, extensive tests for the Ising \cite{r5,tamayo,r12}
and XY \cite{r13,r14} models clearly showed that the single-cluster
variant is the superior algorithm. In our study of the 3D Heisenberg
model we have therefore chosen the single-cluster update algorithm \cite{ours}.
The set-up of our simulations is described in some detail in Sec.~2.

There are of course many sources for a comparison with simulations
based on the standard local heatbath \cite{r15} or Metropolis \cite{r16,r17}
algorithm. In a recent series
of papers Peczak {\em et al.\/} \cite{r18,r19} used the latter algorithm
combined with histogram reweighting techniques \cite{oldpaper,r20,r21,r22} to
estimate the critical coupling and the critical exponents of the
3D Heisenberg model from a
FSS analysis of simulations on simple cubic lattices \cite{r18} and
investigated also its (exponential) autocorrelation time $\tau_0$ \cite{r19}.
Their result, $\tau_0 \propto L^z$, with dynamical
critical  exponent $z = 1.94(6)$, shows that $\tau_0$ is rapidly
increasing
with the linear lattice size $L$ and explains why they could not go to
systems larger than $24^3$. Since, as expected, the autocorrelation time
for the single-cluster algorithm turns out to be almost independent of $L$
we could study much larger lattices of size up to $48^3$ in reasonable
computer time.
Our results described in Sec.~3 provide evidence that the
asymptotic FSS region is indeed reached quite early or, in other
words, that the amplitudes of correction terms are very small.
The present study thus significantly reduces the danger of systematic
errors in the MC estimates of critical exponents from FSS analyses.

In Sec.~4 we report another set of high-precision simulations, done this
time in the
high-temperature phase. There we use variance reduced ``cluster
estimators'' \cite{r10a}
for the spatial correlation length, $\xi$, and the
susceptibility, $\chi$. By going to very large systems with up to
$100^3$ spins we have made sure that these data have only negligible
finite-size corrections well below the statistical errors of about $0.2\%$.
Least-square fits of the critical divergencies of $\xi$ and $\chi$ to the
well-known power-laws yield independent estimates for
the critical coupling and the critical exponents $\nu$ and $\gamma$.
Obviously this is another useful check on residual systematic errors.
Finally, in Sec.~5, we briefly discuss and summarize our main results.
%
            \section{The model and simulation techniques}
%
Let us start with a brief description of the model and some remarks
on the simulation techniques we have used.
The partition function of the Heisenberg model is given by
\begin{equation}
Z = \prod_{\BF{x}} \left[ \int
\frac{d\phi(\BF{x})d\cos\theta(\BF{x})}{4\pi} \right] e^{-\beta E},
\label{eq:1}
\end{equation}
where $\beta \equiv J/k_BT$ is the (reduced) inverse temperature
and the energy is
\begin{equation}
E = \sum_{\BF{x},i} \left[
1-\vec{s}(\BF{x})\vec{s}(\BF{x}+\BF{i}) \right].
\label{eq:2}
\end{equation}
Here $\vec{s} = (\sin \theta \cos \phi, \sin \theta \sin \phi,
\cos \theta)$ are
three-dimensional unit spins at the sites $\BF{x}$ of a
simple cubic lattice of size $V \equiv L^3$, and $\BF{i}$ are unit steps
in the three coordinate directions. We always employ periodic
boundary conditions.
\subsection{Algorithm}
For the closely
related 3D Ising and XY models it has been shown
\cite{r5,r12,r13,r14} that the single-cluster update is the fastest MC
algorithm available. We have therefore chosen this variant for our study of
the 3D Heisenberg model. One update in the single-cluster
variant consists of choosing a random mirror plane and a random
site, which is the starting point for
growing a cluster of reflected spins. The
size and shape of the cluster is controlled by a Metropolis
like accept/reject criterion
satisfying detailed balance \cite{r4,r8}. Compared with the
multiple-cluster algorithm this variant is technically
somewhat simpler to implement and, more importantly,
in three dimensions numerically more efficient. The reason is that,
on the average, larger clusters are moved.

To test the performance of the algorithm we have  recorded autocorrelation
functions, $A(k) = \rho(k)/\rho(0)$, with
\begin{equation}
\rho(k) = \langle O_i O_{i+k}\rangle  - \langle O_i\rangle ^2,
\label{eq:7}
\end{equation}
and $O_i$ denoting the i-th measurement of an observable.
We have focussed on the integrated autocorrelation time,
$ \tau \equiv \frac{1}{2} + \sum_{k=1}^{\infty} A(k)$,
which describes the enhancement of the statistical error,
$\epsilon = \sqrt{\sigma^2/N} \sqrt{2\tau}$, for the mean value over a
sample of $N$ correlated
measurements of variance $\sigma^2$. The infinite sum was always
self-consistently cut off at $k_{\rm max} \approx 6 \tau$
\cite{madras,book1,book2}.

Recall that for the single-cluster update some
care is necessary in defining the unit of time, since in
each update step only a relatively small fraction of the spins
is moved, depending on temperature and lattice size. More precisely,
our results show that near $\beta_c$ and for all lattice sizes, the
average cluster size, $\langle |C| \rangle$, is proportional to the
susceptibility,
\begin{equation}
\langle |C| \rangle \approx 0.75 \bar{\chi}
\label{eq:9}
\end{equation}
(with roughly the same constant as in two dimensions \cite{r10a}), where
$\bar{\chi} = \chi/\beta = V \langle \vec{m}^2 \rangle$,
$\vec{m} = \frac {1}{V} \sum_{\BF{x}}
\vec{s}(\BF{x})$.
At $\beta_c$, the susceptibility behaves like $\chi \propto
L^{\gamma/\nu}=L^{2-\eta}$ (with very small $\eta \approx
0.04$), so that with increasing lattice size the fraction
of moved spins in each update step decreases like
$\langle |C| \rangle/V \propto L^{-(1+\eta)}$, i.e., roughly $\propto 1/L$.
Since the CPU time needed for a single-cluster
update is approximately proportional to the number of moved spins,
it is
convenient
to use $N_0 \equiv V/\langle |C| \rangle \propto
L^{1+\eta}$ single update steps as unit of time. This is
then directly comparable with other schemes that
attempt moves for all spins in one update step. All our
autocorrelation times refer to this unit of time (Metropolis-equivalent
unit) which can always be achieved by a rescaling of the time variable.

For the
susceptibility we typically find $\tau \approx 1.5 - 2.0$.
The values of $\tau$ for each simulation are given in Table~1.
Already our rough estimate of $\tau$ shows that for large system
sizes the single-cluster update clearly outperforms the Metropolis
algorithm, for which the exponential autocorrelation time $\tau_0$ has
recently been determined \cite{r19} to be $\tau_0 = a L^z$, with
amplitude\footnote{
This can be read  off from Fig.~2 in ref.~\cite{r19}.}
$a \approx
3.76$ and dynamical critical exponent $z = 1.94(6)$. For our largest
lattice size $L$=48 this implies a reduction of the autocorrelation time
by about three orders of magnitude.
A more
detailed study of the autocorrelations is in preparation.

To extract the critical exponents we have performed two sets of simulations;
one in the critical region where the correlation length of the infinite
system is much larger than the linear system size, $\xi_\infty \gg L$,
and the other in the high-temperature phase, making sure that $\xi_\infty \ll
L$
to reduce finite-size corrections as much as possible. The quantities we have
measured are the internal energy, specific heat, correlation length, average
cluster size,
susceptibility, and higher order moments. In the critical region
this can be done most efficiently by combining the single-cluster update
with multi-histogram sampling techniques,
and in the high-temperature phase by using
improved estimators for measurements.
\subsection{Improved estimators}
An improved ``cluster estimator'' for the spin-spin correlation function
in the high-temperature phase, $G(\BF{x}-\BF{x}') \equiv \langle
\vec{s}(\BF{x}) \cdot  \vec{s}(\BF{x}') \rangle$, is given by
\cite{r10a} \begin{equation}
\tilde{G}(\BF{x}-\BF{x}') = 3 \frac{V}{|C|} \vec{r} \cdot \vec{s}(\BF{x}) \,\,
\vec{r} \cdot \vec{s}(\BF{x}') \Theta_C(\BF{x}) \Theta_C(\BF{x}'),
\label{eq:3}
\end{equation}
where $\vec{r}$ is the normal of the mirror plane used in the construction
of the cluster of size $|C|$ and $\Theta_C(\BF{x})$ is its characteristic
function (=1 if $\BF{x} \in C$ and 0 otherwise). For the Fourier transform,
$\hat{G}(\BF{k}) = \sum_{\BF x} G(\BF{x}) \exp(-i\BF{k}\BF{x})$, this
implies the improved estimator
\begin{equation}
\tilde{\hat{G}}(\BF{k}) = \frac{3}{|C|} \left[ \left( \sum_{\BF{x} \in C}
\vec{r} \cdot \vec{s}(\BF{x}) \cos \BF{k} \BF{x} \right)^2 +
\left( \sum_{\BF{x} \in C}
\vec{r} \cdot \vec{s}(\BF{x}) \sin \BF{k} \BF{x} \right)^2 \right],
\label{eq:4}
\end{equation}
which, for $\BF{k} = \BF{0}$, reduces to an improved estimator for the
susceptibility $\bar{\chi}$ in the high-temperature phase,
\begin{equation}
\tilde{\hat{G}}(\BF{0}) = \tilde{\bar{\chi}} = \frac{3}{|C|}
\left( \sum_{\BF{x}
\in C} \vec{r} \cdot \vec{s}(\BF{x}) \right)^2.
\label{eq:5}
\end{equation}
Note that we follow in the definition of the susceptibility $\bar {\chi}$
in the high temperature section
the standard convention of omitting the $\beta$-factor. From the theoretically
expected small $\BF{k}$ behavior of the inverse  Fourier transform,
\begin{equation}
\hat{G}(\BF{k})^{-1} = c \left[ \sum_{i=1}^3 2 (1 - \cos k_i) + (1/\xi)^2
\right]
\approx c \left[ \BF{k}^2 + (1/\xi)^2 \right], \label{eq:6} \end{equation}
where $c$ is a constant and $k_i = (2\pi/L)n_i$, $n_i = 1,\dots,L$,
one may extract the
correlation length $\xi$ by measuring $\hat{G}$ for a few long-wavelength
modes and performing least-square fits to (\ref{eq:6}). In our simulations
we have measured $\hat{G}(\BF{k})$ for $\BF{n} =$ (0,0,0), (1,0,0), (1,1,0),
(1,1,1), (2,0,0), and (2,1,0) (see Sec.~4).
It is well known that by means of the estimators (\ref{eq:3})-(\ref{eq:5})
a significant reduction
of variance can only be expected outside the FSS region
where the average cluster size is small compared with the volume of the system.
\subsection{Histogram techniques}
Even though histogram reweighting techniques have been known since long
\cite{oldpaper}, they have gained increasing popularity as a practical tool
only quite recently \cite{r20}. The best performance is achieved {\em near}
criticality, and in this sense the histogram reweighting
technique is complementary to the use of improved estimators. It is a quite
general technique of data analysis based on the simple idea of recording
whole distribution functions, and not only their first few
moments (e.g., the average energy and specific heat), as is usually done.
The energy distribution ${\cal P}_{\beta_0}(E)$ (normalized to unit area) at
inverse temperature $\beta_0$ can be written as
\begin{equation}
{\cal P}_{\beta_0}(E) = \rho(E) e ^{-\beta_0 E} / Z(\beta_0),
\label{eq:13}
\end{equation}
where $\rho(E)$ is the density of states with energy $E$.
It is
then easy to see that an expectation value $\langle f(E) \rangle$
can in principle be calculated for any $\beta$ from
\begin{equation}
\langle f(E) \rangle(\beta) = \frac{\int_0^{\infty} dE f(E) {\cal P}_{\beta_0}
(E) e^{-(\beta-\beta_0)E}}{\int_0^{\infty} dE {\cal P}_{\beta_0}
(E) e^{-(\beta-\beta_0)E}}.
\label{eq:14}
\end{equation}
To keep the notation short we have suppressed the lattice-size dependence
of ${\cal P}_{\beta_0}(E)$.

In practice the continuous energy has to be discretized
into bins of size $\Delta E$, and one measures the associated histogram
$P_{\beta_0}(E) = {\cal P}_{\beta_0}(E)\Delta E$. Since the wings
of $P_{\beta_0}(E)$ have large  statistical errors, one expects eq.
(\ref{eq:14}) (or its obvious discrete modification) to give reliable
results only for $\beta$ near $\beta_0$. If
$\beta_0$ is near  criticality, the distribution is relatively broad
and the method
works best. In this case reliable estimates from (\ref{eq:14})
can be expected for $\beta$ values in an interval around
$\beta_0$ of width $\propto L^{-1/\nu}$, i.e., just in the
FSS region. The Fig.~1(a) shows three typical energy
histograms measured at slightly different temperatures near the critical point
for the lattice size $L$=48.

The information stored in $P_{\beta_0}(E)$ is not yet sufficient to calculate
also
the magnetic moments
$\langle m^k \rangle (\beta)$ with
$m = |\vec{m}|$
as function of $\beta$ from a single simulation at $\beta_0$.
Conceptually, the simplest way to do so
is to record the two-dimensional histogram
$P_{\beta_0}(E,M)$, where $M = mV$ is the total magnetization (in
general we will denote by small letters quantities per site, and the associated
total quantities by the corresponding capital letters). Since disk space
limitations prevented us from doing that\footnote{As well as from storing the
full time series of $E$ and $M$, which conceptually would be even simpler.}
we have measured instead the ``microcanonical averages''
\begin{equation}
\langle\!\langle m^k
\rangle\!\rangle(E) = \sum_M P_{\beta_0}(E,M) m^k /P_{\beta_0}(E),
\label{eq:14a}
\end{equation}
where we have used the trivial relation
$\sum_M P_{\beta_0}(E,M) = P_{\beta_0}(E)$. In practice this can be done simply
by accumulating the measurements of $m^k$ in different slots or bins
according to the energy of the configuration and normalizing at the end by the
total number of hits of each energy bin. Clearly, once
$\langle\!\langle m^k \rangle\!\rangle(E)$ is determined, this is
a special case of $f(E)$ in eq.(\ref{eq:14}), so that
\begin{equation}
\langle m^k \rangle(\beta) = \frac{
\sum_E \langle\!\langle m^k \rangle\!\rangle(E) P_{\beta_0}(E)
e^{-(\beta - \beta_0)E}}
{\sum_E P_{\beta_0}(E) e^{-(\beta - \beta_0)E}} .
\label{eq:14b}
\end{equation}
Similar to $\rho(E)$ in (\ref{eq:13}), theoretically the microcanonical
averages
$\langle\!\langle m^k \rangle\!\rangle(E) $ do not depend
on the temperature at which the simulation is performed. Due to the limited
statistics in the wings of $P_{\beta_0}(E)$, however, there is only a finite
range around $E_0 \equiv \langle E \rangle (\beta_0)$ where
one can expect reasonable results for
$\langle\!\langle m^k \rangle\!\rangle(E) $. Outside of this
range it simply can happen (and  does happen) that there are no events to be
averaged. This is illustrated in Fig.~1(b), where
$\langle\!\langle m \rangle\!\rangle(E) $ is plotted versus energy
as obtained from the three runs
which gave the energy histograms in Fig.~1(a).
We see that the function looks smooth only in the range where the
corresponding energy histogram in Fig.~1(a) has enough statistics.

To take full advantage of the histogram reweighting technique we have performed
for each $L$ typically three\footnote{
Due to technical reasons we had four simulations for the lattices
$L$=24 and $L$=48, compare also Table~1.}
simulations at slightly different inverse temperatures $\beta_i$. Using
histogram reweighting and jackknife-blocking \cite{r27} we computed the
$\beta$-dependence of the expectation values for all interesting thermodynamic
observables  $O_i \equiv O_L^{(\beta_i)}(\beta)$ plus the associated
error $\Delta O_i$.  To obtain a single expression $O \equiv O_L(\beta)$
we then combined the values
$O_i$ numerically according to the formula
\begin {equation}
O = \left( \frac{O_1}{\left( \Delta O_1\right)^2} +
           \frac{O_2}{\left( \Delta O_2\right)^2} +
           \frac{O_3}{\left( \Delta O_3\right)^2} \right)
    \left(\Delta O \right)^2 ,
\label{eq:30}
\end {equation}
where $\Delta O$ is given by
\begin {equation}
\frac{1}{\left( \Delta O \right)^2} =
                                      \frac{1}{\left( \Delta O_1\right)^2} +
                                      \frac{1}{\left( \Delta O_2\right)^2} +
                                      \frac{1}{\left( \Delta O_3\right)^2} .
\end {equation}
This expression minimizes the relative error $\Delta O/O$.
The reweighting range of each simulation, i.e., that range in
which the energy histogram has enough statistics to allow for (\ref{eq:14})
to be valid, was determined by the energy values at which the histogram
had decreased to a third of its maximum value \cite{alves}. From the energy
range one can then deduce a corresponding $\beta$-range. Only the values inside
this $\beta$-window were used for the optimized combination (\ref{eq:30}), the
contributions of the other values were suppressed by giving them zero weights.

We also implemented the
optimized histogram combination discussed in ref.~\cite{r21} to
cross-check the results obtained by (\ref{eq:30}) for the specific heat.
The Fig.~1(c) shows the weights for the optimal combination of the
primary energy
histograms according to the prescription of ref.~\cite{r21}, i.e., which gives
the
optimal combination of the three estimates for $N(E) \equiv \rho (E) \Delta E$.
Both
methods gave comparable results within the statistical errors. We preferred our
optimization procedure over the optimized histogram addition because it
is simpler
to apply to quantities involving constant energy averages such as $\langle\!
\langle m \rangle\!\rangle(E)$, and, more importantly,  minimizes
the error on each observable of interest separately.
%
   \section{Results at criticality and finite-size scaling analysis}
%
We investigated in our MC study simple cubic lattices of volume
$V=L^3$, where
$L$=12, 16, 20, 24, 32, 40, and 48. For each $L$ we have made at least
three simulations at three different temperatures compiled
in Table~1. For all
$L$ we took $\beta=0.6929$,
because this is
the critical inverse temperature found in the recent study of
Peczak {\em et al.}
\cite{r18}. We analyzed this run to locate a first estimate for the
temperatures
of the maxima of the specific heat and the susceptibility for each $L$, and
used
those two temperatures for our other two simulations. This choice has the
advantage
that both locations of the maxima scale approximately like $L^{-1/\nu}$
which is also the region after which the energy distribution $P_\beta(E)$
has decreased by roughly the same factor for all $L$. We can therefore
expect to have always enough overlap to ensure a safe reweighting of
our three histograms into the $\beta$-region of interest. A further advantage
is
that the two maxima approach $T_c$ from different sides.

For each configuration we recorded the energy histogram $P_\beta(E)$, using
90000 bins to discretize the continuous energy range $0 \le E \le 3V$.
We have checked that this binning is fine enough to ensure negligibly small
discretization errors.
In addition we
recorded the microcanonical
averages of $\langle\!\langle m \rangle\!\rangle(E)$,
$\langle\!\langle m^2 \rangle\!\rangle(E)$, and
$\langle\!\langle m^4 \rangle\!\rangle(E)$.
These histograms provided us with all
necessary information to calculate all thermodynamic quantities of interest.
Every 10000 measurement steps we recorded
a copy of each histogram, so that we were able to compute errors by
standard jackknife-blocking \cite{r27}.
%
            \subsection{Binder parameter $U_L(\beta)$}
%
We have used the histogram reweighting technique to
find the $\beta$-dependence of the Binder parameter
\cite{bi82},
\begin{equation}
U_L(\beta) = 1 - \frac{1}{3}\frac{\langle m^4\rangle }{\langle m^2\rangle ^2}.
\label{eq:15}
\end{equation}
To obtain a single curve $U_L(\beta)$ for each lattice size $L$ from the three
simulations at temperatures
compiled in Table~1 we used our optimized combination of eq.~(\ref{eq:30}).
In (\ref{eq:15})
we adapt the usual  normalization convention, although the factor 3 is only
really
motivated for the Ising ($O(1)$) model. For general $O(n)$  models it is easy
to show that
in the high-temperature limit Gaussian fluctuations around $\vec{m}=0$ lead
to $U_L
\rightarrow  2(n-1)/3n$. For $n=1$ this gives a zero reference point, but
for the Heisenberg model with $n=3$ we get the quite
arbitrary looking limit $U_L \rightarrow 4/9$. In three dimensions,
we are for low temperatures always in the magnetized phase and hence have
for all $n$ trivially $U_L \rightarrow 2/3$.

It is well known \cite{bi82}
that the $U_L(\beta)$ curves for different $L$ cross around
$(\beta_c,U^{*})$ with slopes $\propto L^{1/\nu}$, apart from
confluent corrections explaining small systematic deviations. This
allows an almost unbiased estimate of $\beta_c$, the
critical exponent $\nu$, and $U^{*}$.
Field theoretical expansions in
$\sqrt{\epsilon} \equiv \sqrt{4-D}$ predict \cite{br85}
\begin{equation}
U^* = 0.59684\dots .
\label{eq:15b}
\end{equation}
This follows from the expansion
\begin{eqnarray}
\frac{\langle m^4 \rangle}{\langle m^2
\rangle^2} &=& \frac{3}{4}\frac{\Gamma^2(3/4)}{\Gamma^2(5/4)}
\Bigg[ 1 - x_0 \sqrt{6} \left(
\frac{\Gamma(9/4)}{\Gamma(7/4)} +
\frac{\Gamma(5/4)}{\Gamma(3/4)}
-2\frac{\Gamma(7/4)}{\Gamma(5/4)} \right) \nonumber \\
&+& 6 x_0^2 \left(
\frac{\Gamma(9/4)\Gamma(5/4)}{\Gamma(7/4)\Gamma(3/4)}
+3 \frac{\Gamma^2(7/4)}{\Gamma^2(5/4)} - 4 \right) \Bigg]
\label{eq:15c} \\
&=& \frac{12}{R^2}\left[ 1 - x_0 \sqrt{6} \left(
\frac{2}{3}R - \frac{6}{R} \right) + 6 x_0^2 \left(
\frac{5}{48} R^2 + \frac{27}{R^2} -4 \right) \right],
\nonumber
\end{eqnarray}
where $x_0 = -1.7650848012 \frac{5}{2\sqrt{33}}
\sqrt{\epsilon} = -0.76815456 \, \sqrt{\epsilon}$, and in the
last line we have abbreviated $R \equiv
\Gamma(1/4)/\Gamma(3/4) = 2.95867512\dots$.
Notice also that (\ref{eq:15}) can be rewritten as
\begin{equation}
U_L = \frac{2}{3} - \frac{\sigma^2_{\chi}}{3\chi^2},
\label{eq:15a}
\end{equation}
where $\sigma^2_{\chi} \equiv V^2 \beta^2 \left( \langle m^4 \rangle
 - \langle m^2 \rangle ^2 \right)$ is the
variance of the susceptibility.
In the $T \rightarrow \infty$
limit we get $\sigma^2_{\chi} \rightarrow \frac{2}{3} \chi^2$
and at criticality, inserting the field theory
prediction (\ref{eq:15b}), this implies
$\sigma^2_{\chi} \approx 0.21 \chi^2$.

Figure~2 shows the crossings of the $U_L(\beta)$ curves on a large scale,
which gives a first estimate of ($\beta_c,U^*$).
To extract more precise values of $U^*$ and $\beta_c$ from our data
we used that
the locations of the crossing point $\beta^{\times} \equiv 1/T^{\times}$
of two different curves $U_L(\beta)$ and $U_{L'}(\beta)$ depend on the
scale factor $b=L'/L$,
due to the residual corrections to FSS \cite{bi82}.
To have enough data points
for a straight line fit we used only the crossing points of
the $L$=12 and $L$=16 curves with all the other ones with higher $L'$-value,
which gave us 6 and 5 data
points, respectively.
The two least-square extrapolations in
the plot of $T^{\times}$ vs $1 / \log b$ shown in Fig.~3 are consistent with
each other and gave us the values $\beta_c = 0.69297(9)$ and
$\beta_c = 0.69298(13)$ respectively. Combining the two values we obtained
\begin{equation}
\beta_c = 0.6930 \pm 0.0001.    \label{eq:16}
\end{equation}
The errors on $T^{\times}$ were obtained by using the crossings of
$U_L(\beta) + \Delta U_L(\beta)$
with $U_{L'}(\beta) - \Delta U_{L'}(\beta)$,
where $\Delta U_L$ are the errors on $U_L$ obtained via the jackknife
procedure.
Our result (\ref{eq:16}) is in good agreement with the value given in
ref.~\cite{r18}, $\beta_c = 0.6929(1)$, but significantly higher than
estimates from analyses of high-temperature series expansions
\cite{rf,series2}.

A similar analysis \cite{bi82} was used to determine $U^*$, which resulted in
$U^*=0.62175(35)$
and $U^*=0.62141(75)$, repectively, from which we extract the final estimate
\begin{equation}
U^{*} = 0.6217 \pm 0.0008.      \label{eq:17}
\end{equation}
The theoretical prediction (\ref{eq:15b}) based on the
$\sqrt{\epsilon}$-expansion is about
$4\%$ smaller than this value.
The deviation is somewhat less than
for the XY model where it is about $6\%$ \cite{r13,r14}.

To determine the derivative  ${dU_L}/{d\beta}$, we first used a finite
difference approximation at our estimate of $\beta_c$. But we observed that the
results depended very sensitively on the interval of the
linear approximation, and on the kind of finite difference derivative used,
i.e., backward, forward, or symmetric derivative. We have therefore chosen
another method \cite{r26} to calculate the slope
of $U_L(\beta)$ that is less sensitive to systematic errors. We took the
thermodynamic derivative of $U_L$ with respect to $\beta$, which can
be written as
\begin{eqnarray}
{{dU_L} \over {d\beta }} &=&
{1 \over {3\langle m^2\rangle ^2}} \left\{
{\left\langle {m^4} \right\rangle\langle E\rangle
-2{{\left\langle {m^4} \right\rangle\langle m^2E\rangle } \over
{\left\langle {m^2} \right\rangle} }
+\langle m^4E \rangle
}
\right\}\hfill\cr  &=&(1-U_L)\left\{ {\langle E\rangle -2{{\langle m^2E\rangle
}  \over {\left\langle {m^2} \right\rangle}}+{{\langle m^4E\rangle } \over
{\left\langle {m^4} \right\rangle}}} \right\}. \label{eq:18a}
\end{eqnarray}
The expectation values for $\langle m^kE \rangle(\beta)$ at any
temperature $\beta$ can be calculated from
\begin{equation}
\langle m^kE\rangle (\beta )={{\sum\limits_E {E\langle \langle
m^k\rangle \rangle (E)P_{\beta_0}(E)e^{-(\beta -\beta _0)E}}}  \over
{\sum\limits_E {P_{\beta_0}(E)e^{-(\beta -\beta _0)E}}}},
\end{equation}
generalizing eq.~(\ref{eq:14}).
For each $L$ and $\beta$ we obtained in this way
three estimates of $\frac {dU_L} {d\beta}(\beta)$
that we combined optimally according to the errors of $\frac {dU_L} {d\beta}$.

In Fig.~4 we plot $\frac {dU_L} {d\beta} (\beta_c)$, taken at our estimate of
$\beta_c=0.6930$, versus $L$ on a log-log scale. From the inverse slope
we read off
  \begin{equation}
\nu = 0.704 \pm 0.006, \label{eq:18}
\end{equation}
with a quality factor $Q$=0.61, relying on the linear least-square fit routine
FIT of ref.~\cite{r24}. For comparison, the field theoretical estimates are
$\nu = 0.705(3)$
(resummed perturbation series \cite{zi80}),
$\nu = 0.710(7)$
(resummed $\epsilon$-expansion \cite{zi85}). We noted that the value of $\nu$
derived in this way
was strongly dependent on the temperature at which it was
extracted. It turned out that the systematic errors one gets by choosing
a slightly different value for $\beta_c$ are larger than the statistical
errors. The strong $\beta$-dependence of $\frac {dU_L} {d\beta}$ has its
origin in the large system sizes we used that make the thermodynamic
quantities vary very rapidly with $\beta$. For comparison, for $\beta=0.6929$
($\beta=0.6931$) we obtained $\nu=0.696(6)$ with $Q$=0.76 ($\nu=0.712(6)$ with
$Q$=0.49); see also Fig.~8 below.
%
           \subsection{Magnetization and susceptibility}
%
To extract the ratio of critical exponents $\beta/\nu$ we used that
the magnetization at $\beta_c$
should scale for sufficiently large $L$ like \cite{barber}
\begin{equation}
\langle m \rangle \propto L^{-\beta /\nu }.
\label{eq:18c}
\end{equation}
The slope of the straight line in the log-log plot shown in Fig.~5
of $\langle m \rangle$ vs $L$ at $\beta_c = 0.6930$ gives a value of
\begin{equation}
\beta/\nu=0.514\pm 0.001,
\label{eq:18d}
\end{equation}
with a $Q$-value of 0.68 of our linear
least-square fit routine. We noted again that the systematic error of
choosing an incorrect
$\beta_c$ was larger than the statistical error coming from the fit.
For example, for fits of
$\langle m \rangle$ at the two inverse temperatures $\beta = 0.6929$
and $\beta=0.6931$ we
obtained $\beta/\nu=0.519(1)$ and $\beta/\nu=0.509(1)$ with
values $Q=0.30$ and $Q=0.31$, respectively. We take the significant lower
$Q$-values as supporting our estimate of $\beta_c$ (compare also Fig.~8 and
the discussion of the $Q$-values later in this section). Using our
estimate of $\nu = 0.704(6)$ we get for the critical exponent of the
magnetization $\beta = 0.362(4)$.

On finite lattices an
estimator for the magnetic susceptibility per spin is given by
\begin{equation}
\chi^{\rm c} = V\beta  (\langle m^2\rangle  - \langle m\rangle ^2),
\end{equation}
where the superscript c stands short for ``connected''. In the
high-temperature phase the true magnetization vanishes,
$\langle \vec{m} \rangle = 0$, and one may also use
\begin{equation}
\chi = V\beta  \langle m^2\rangle   \mbox{~~~~for~~} \beta <  \beta_c .
\end{equation}
Since both expressions should scale at criticality like $L^{\gamma /
\nu}= L^{2-\eta}$ a straight line fit in a double logarithmic plot of
$\chi / L^2$
versus $L$ gives an estimate of $\eta$. We preferred to visualize $\eta$
as opposed to
$\gamma/\nu$ because $\eta$ is the more sensitive parameter.
The results for different choices of $\beta_c$ are shown in Fig.~6. We see that
our estimate of $\beta_c$ in (\ref{eq:16}) seems to be very reasonable, which
is
also supported by the fact that it results in the fit with the
highest $Q$-value.
The fact that the curves with $\beta > \beta_c$
($\beta < \beta_c$) bend upwards (downwards) is easily understood by
recalling that in the low-temperature phase $\chi$ scales with the
volume $V = L^3$, while in the high-temperature phase $\chi$ eventually
saturates at a finite value for any $L$. Close to criticality, as long
as the scaling variable $x \equiv (1-\beta/\beta_c) L^{1/\nu}$ satisfies
$|x| \ll 1$ and $L \gg 1$, we expect from FSS that the deviations from
$-\eta \log L + c_0$ should be simply given by $c_1 x$, where $c_0$ and
$c_1$ are approximately constant (apart from small confluent corrections
$\propto L^{-\omega}$). For all data points in Fig.~6 we have $|x| < 0.18$
and the linear scaling with $(1-\beta/\beta_c)$ is obviously satisfied.
Moreover, a closer look shows that also the scaling with $L^{1/\nu}$ is
very well confirmed (using our estimate of $\nu = 0.704$ the deviations from
the straight line fit of the curves at $\beta \ne \beta_c$ for $L=32,40$
and $48$ should be stronger than those for $L=24$ by factors of about
1.5, 2.1 and 2.7, respectively). In fact, in a scaling plot all curves
would fall on top of each other.
The theoretical predictions for $\eta$
are 0.040(3) (resummed
$\epsilon$-expansion) and 0.033(4) (resummed perturbation series),
and inserting (\ref{eq:18d}) in the scaling relation $\eta = 2\beta/\nu -1$
gives $\eta = 0.028(2)$. Our direct
estimates of $\eta$ for
$\beta=0.6930 \pm 0.0001$ are collected in Table~2. It is also interesting
to note that
$\eta$ gets closer to the theoretical values for $\beta<\beta_c$ for $\chi$,
and for
$\beta>\beta_c$ for $\chi^{\rm c}$, hence just in the regions in which both
expressions are naturally used. Another check on $\eta$ can be
obtained from the
maxima of $\chi^{\rm c}$, that should obey the same scaling law as
$\chi^{\rm c}$ and
$\chi$ at $\beta_c$. The results are compiled in Table~2. Our three estimates
for $\eta$
barely agree in the 2$\sigma$ range, but this is mainly due to the low
estimate of $\eta$
coming from the $\chi^{\rm c}$-fit. This behavior of the  $\chi^{\rm c}$
scaling law is in
agreement with the findings of refs.~\cite{r18,r23}, who also observed
that the $\chi^{\rm c}$-fit results in a noticeably lower $\eta$ value.
As our final estimate we take the result from the best fit of $\chi$ at
$\beta_c = 0.6930$, yielding
\begin{equation}
\eta = 0.0271 \pm 0.0017.
\end{equation}
This in turn implies $\gamma/\nu = 2 -\eta = 1.9729(17)$ and, using again
our value of $\nu$, $\gamma = 1.389(14)$.

{}From the temperature locations $T_{\chi^{\rm c}_{\rm max}}$ of the
(connected)
susceptibility maxima one
can get another estimate of $\beta_c$ by using the scaling prediction
 $T_{{\chi}_{\rm max}^{\rm
c}}=T_c + aL^{-1/\nu}+\dots$. Using our previously determined value
 of $\nu =0.704$ we get
from the linear fit shown in Fig.~7 the estimate $\beta_c
=0.6930(3)$, with $Q$=1, which is in excellent agreement
with our result (\ref{eq:16})
for $\beta_c$ obtained from the $U_L$ crossing points. The data for the fit is
compiled in Table~3.

When we did our least-square fitting of quantities extracted at a
particular $\beta$, we noticed that
the $Q$-value can also be used as a very sensitive measure to obtain an
estimate of the critical
temperature, whose error is comparable to the one obtained
from the fits to the $U_L$ crossing points. Obviously if one extracts
data at
temperatures away from $\beta_c$  the subleading corrections to the scaling
laws become more
important and worsen the fit substantially, especially if large lattices are
used. In Fig.~8 we
plot the quality factor $Q$ obtained from log-log linear least-square fits of
the quantities
$\frac {dU_L} {d\beta} (\beta), \langle m \rangle (\beta), \chi (\beta)$,
and $\chi^{\rm c} (\beta)$ vs the lattice size $L$, extracted at different
inverse temperatures. The most
striking results come from the $Q(\beta)$-curves of $\langle m \rangle$
and $\chi$, that seem to have the
most severe $\beta$-dependence. Their peak locations are exactly at
$\beta = 0.6930$ with an estimated error of $\pm 0.0001$, whereas the
$Q(\beta)$-curves of
$\frac {dU_L}{d\beta} (\beta)$ and $\chi^{\rm c}(\beta)$
do not peak so sharply at some $\beta$ and
are asymmetric with respect to their peak location, falling off very
slowly on the high-temperature side. The slow variation of the $Q$-value
of the $\frac {dU_L}{d\beta} (\beta)$-fits indicates that Binder's method to
determine
$\nu$ is less sensitive to the choice of $\beta_c$ than the FSS method for
$\beta /
\nu$ and $\gamma / \nu$, as is expected on theoretical grounds \cite{bi82}.
This is reflected in our data for $\nu$, where the systematic errors for
different
choices of the critical coupling are of the order of the statistical errors,
whereas
for $\beta / \nu$ they are five times as big.
%
           \subsection{Energy and specific heat}
%
Assuming hyperscaling, $\alpha = 2 - D \nu$,  to be valid and inserting
our estimate $\nu = 0.704(6)$, we get a negative value
for the critical exponent of the specific heat, $\alpha = -0.112(18)$.
This implies a cusp like singularity at $T_c$,
but no divergence at all. Although at first sight this sounds
numerically very convenient, it turns out that the specific heat is the
hardest observable to analyze. On one hand it is quite
easy to get the peak height with good precision due to the smoothness of
the specific
heat. On the other hand, however, the scaling behavior is
quite unclear for $\alpha < 0$, since one has to
deal with large, non-universal background terms. For the peak location,
$T_{C_{\rm max}}(L)$, the scaling behavior is of the usual form
$T_{C_{\rm max}}(L) \approx T_C +
aL^{-1/\nu}$, but now the smoothness of the peaks makes it
numerically difficult to determine
$T_{C_{\rm max}}(L)$ with high precision.

We calculated the specific heat in two different ways, once from the energy
fluctuations as
\begin{equation}
C = V \beta^2 (\langle e^2\rangle  - \langle e\rangle ^2),
\label{eq:20c}
\end{equation}
and
the other time as a finite difference derivative approximating $C=de/dT$.
Both definitions resulted in similar curves, and this time the choice of
the finite difference scheme did not affect the results. Because the curves
of definition (\ref{eq:20c}) were smoother in appearance, we decided to use
solely the first definition to extract our scaling information. The maximum
of the specific heat should scale as
\begin{equation}
C_{\rm max}(L) \approx C_{\rm max}^{\rm
reg} - aL^{\alpha/\nu}, \label{eq:20d}
\end{equation}
where $C_{\rm max}^{\rm reg}$ is a regular background term.
Using the routine MRQMIN of
ref.~\cite{r24} we obtained from a three-parameter fit the values
$C_{\rm max}^{\rm reg}= 4.17(97)$, $a=3.98(17)$,
$\alpha / \nu = -0.33(22)$, with a $Q$-value of $Q$=0.69.
Using our estimate of $\nu = 0.704(6)$, this results
in $\alpha = -0.23(16)$. This estimate is clearly larger
than the one obtained
through hyperscaling but, due to its large error bar, it is still consistent.
By imposing the hyperscaling
value of $\alpha/\nu = 2/\nu - 3 = -0.159$, we also did a
linear fit of $C_{\rm max}(L)$ vs $L^{\alpha/\nu}$ that gave us
$C_{\rm max}^{\rm reg}= 5.79(12) $, and
$a= 5.00(19) $ with\footnote{
Although the $Q$-value for the 2-parameter fit is better, its total
$\chi^2$(=2.9) is larger than that of the 3-parameter fit (=2.2).}
$Q = 0.71$. Both curves fit the data
almost equally good as can be inspected in Fig.~9.

The fit of $T_{C_{\rm max}}(L)$ versus $L^{-1/\nu}$ is shown is Fig.~7 .
Assuming $\nu=0.704$, our linear fit routine gave
$\beta_c=0.6925(9)$ with $Q$=0.80. Due to the
large error bars on $T_{C_{\rm max}}$ this estimate has the largest statistical
scatter, but is still in agreement with our previous estimates of $\beta_c$.
The
data for the fit can be found in Table~3.
%
             \section{Results in the high-temperature phase}
%
Let us now turn to the analysis of our second set of simulations which were
performed in the high-temperature
phase. Here we have mainly concentrated on measurements of the spatial
correlation length,
$\xi$, and the susceptibility, $\bar {\chi}$, using the improved estimators of
Sec.~2.2. From least-square fits to the critical
divergencies of $\xi$
and $\bar {\chi}$ as function of $T$ or $\beta$ we can get independent
estimates of the critical coupling $\beta_c$ and of the critical exponents
$\nu$ and $\gamma$. This of course requires to keep finite-size corrections
as small as possible. We therefore have chosen the linear size of the lattices
to satisfy the condition $L \ge 8 \xi$, which proved in related
studies \cite{r9,r9a} to
be a safe condition. In a few cases we have checked this once again by
performing test runs at smaller $L \approx 5-7\xi$. As a result, to avoid the
most severe finite-size corrections, also for this model it would probably be
sufficient to choose $L \approx 6 \xi$.

Our final data set consists of 18
points between $\beta=0.650$ and $\beta=0.686$ (corresponding to correlation
lengths $\xi=3,\dots,12$, see below), using lattices of sizes between
$32^3$ and $100^3$. The simulation parameters together with the statistics and
the results are compiled in Table~4. The statistics is given in terms of $N$,
the number
of measurements of the standard estimators which, on the average, were taken
after $f \times V$ spins are flipped. The number of Metropolis equivalent
sweeps is then simply $f \times N$. The average cluster size
$\langle |C| \rangle$ is given in terms of the ratio
$\langle |C| \rangle / \bar {\chi}_{\rm imp}$ which is only very slowly
varying
over a wide temperature range. The column labeled with $C$ shows the specific
heat calculated from the energy fluctuations.
For comparison we give for the susceptibility the averages over
standard as well as over improved estimators. We see
that the errors on $\bar {\chi}_{\rm imp}$ are smaller by about a factor of
$2-3$.
It should be mentioned that the errors on $\bar {\chi}$ could be somewhat
reduced by doing measurements more frequently (i.e., by choosing smaller values
of $f$). The reason is that the (integrated) autocorrelation time for
$\bar {\chi}$
is still very small on the time scale at which the measurements are
taken ($\approx 0.8$, translating in Metropolis equivalent units to
$0.8 \times f \approx 0.1$). On the other hand it should be kept in mind that
each standard measurement takes ${\cal O} (V)$ operations, so that too small a
factor $f$ would slow down the simulation considerably. The correlation length
in the last column was extracted from fits to the inverse Fourier transform
of the spatial correlation function as described in Sec.~2.2. We always used
the lattice momentum squared, $\sum_{i=1}^3 2(1-\cos k_i)$,
$k_i = (2\pi/L) n_i$, as independent variable.  In Fig.~10 this
procedure is illustrated for our largest lattice of size $100^3$. We have
checked that within error bars the estimates for $\xi_{\rm imp}$ do
not depend on how many Fourier coefficients we use in the fits. This
observation
is also supported by the very reasonable values of the goodness-of-fit
parameter for
all fits.  Notice that even the simplest expression,
$\xi_{\rm imp} = (\hat{G}(\BF{0})/\hat{G}(\BF{1}) -1)^{1/2}/2\sin(\pi/L)$
(involving {\em no} fit at all), can be used. Our final results for
$\xi_{\rm imp}$ in Table~4 are from fits using all six possible $\BF{k}$ values
up to $\BF{k} = (2\pi/L) (2,1,0)$, since the error estimates trivially decrease
with the number of points taken into account.

In what follows we describe the least-square fit techniques employed to
determine the critical coupling and the exponents $\nu$ and $\gamma$ from
the raw data in Table~4. Although we are quite confident in the accuracy of
the correlation length results, we shall first consider the susceptibility data
which, \'a priori, is more reliable since no intermediate analyses are
involved.
\subsection{Susceptibility}
Starting with the simplest ansatz,
\begin{equation}
\bar {\chi}(\beta) = \bar {\chi}_{0} (1 - \beta/\beta_c)^{-\gamma},
\label{eq:4.1}
\end{equation}
we obtained from a non-linear three-parameter least-square fit to all 18 data
points (using $\bar {\chi}_{\rm imp}$)
\begin{eqnarray}
\beta_c  &=& 0.69294 \pm 0.00003, \label{eq:4.2a} \\
\gamma   &=& 1.391   \pm 0.003,   \label{eq:4.2b} \\
\bar {\chi}_{0} &=& 0.955   \pm 0.006,   \label{eq:4.2c}
\end{eqnarray}
with a chi-squared of $\chi^2 = 7.83$, corresponding to a
goodness-of-fit parameter $Q = 0.93$, which clearly justifies \'a
posteriori the ansatz  (\ref{eq:4.1}). The very precise estimate
for $\beta_c$ is only little lower than our previous value
(\ref{eq:16}), $\beta_c = 0.6930(1)$, derived from  FSS analyses of
the $U_L$ crossing points. The exponent $\gamma$ agrees very well
with the field theoretical estimates,  $\gamma = 1.386(4)$
(resummed perturbation series \cite{zi80}), $\gamma = 1.390(10)$
(resummed $\epsilon$-expansion \cite{zi85}).  The amplitude $\bar
{\chi}_0$ is close to the mean-field value, $\bar {\chi}_0^{\rm MF}
= 1$, and agrees surprisingly well with old analyses of 9-term
high-temperature series (HTS) expansions by Ritchie and Fisher
\cite{rf} (RF) which gave $\bar {\chi}_0^{\rm RF} = 0.96647(5)$
(using $\beta_c^{\rm RF} = 0.6916(2)$ and  $\gamma^{\rm RF} =
1.38(2)$ as input). We could not trace any more recent published
number to compare with. From the HTS analyses in
ref.~\cite{series2}, however, it is straightforward to derive
another comparative value for  $\bar {\chi}_0$.   Expanding the
ansatz for the dominant singularity (\ref{eq:4.1})
into a power series in $\beta$,
inserting the estimates for the
critical parameters given in Table~4 of ref.~\cite{series2},
$\beta_c = 0.6924(2)$, $\gamma = 1.387(4)$ (Pad\'e approximants) and
$\beta_c = 0.6925(1)$, $\gamma = 1.395(5)$ (ratio method), and comparing
with the coefficients of the HTS
expansion \cite{series2,seriescoef}, we obtain
a sequence of
amplitude values that stabilize with increasing order of the expansion at
$\bar {\chi}_0^{\rm HTS} \approx 0.9508$
and $\bar {\chi}_0^{\rm HTS} \approx 0.9326$, respectively.
Moreover, using our values (\ref{eq:4.2a}) and (\ref{eq:4.2b}) for $\beta_c$
and $\gamma$, we obtain by
the same procedure a very smooth sequence for $\bar {\chi}_0$ yielding
asymptotically
$\bar {\chi}_0 = 0.9501$. The smoothness indicates that the simplest ansatz
(\ref{eq:4.1})
together with our estimates (\ref{eq:4.2a})-(\ref{eq:4.2c}) represents an
excellent extrapolation of the HTS expansion, whose
quality is comparable (if not better) with the results given in
ref.~\cite{series2}.

As a further self-consistency check of the range of
validity of the ansatz (\ref{eq:4.1}) we tested for the asymptotic scaling
region by successively discarding more and more data points with small
correlation length. As is demonstrated in Fig.~11 for $\beta_c$,
only a weak downward trend is observable which, compared to the error bars,
is hardly significant.
We can thus
conclude that the ansatz (\ref{eq:4.1}) passes all usual self-consistency
checks
and that the estimates (\ref{eq:4.2a})-(\ref{eq:4.2c}) should thus be reliable.

It is of course surprising that the simple ansatz (\ref{eq:4.1}) works even
for data points with correlation lengths as small as $\xi \approx 3$.
Theoretically
we would have expected to see {\em confluent} correction terms of the type
$\bar {\chi}_{\rm conf} (1 - \beta/\beta_c)^{-\gamma + \Delta_1},$
where $\Delta_1 = \omega \nu \approx 0.55$ \cite{zi80,zi85}\footnote{
The precise values are $\omega = 0.78(2)$, $\Delta_1 = 0.55(3)$ \cite{zi80};
$\omega = 0.79(4)$, $\Delta_1 = 0.56(4)$ \cite{zi85}.}
is the confluent correction exponent \cite{wegner},
and {\em analytic} correction terms of the Darboux type,
$\bar {\chi}_{\rm anal} (1 - \beta/\beta_c)^{-\gamma + 1}$, leading to
the more general ansatz
\begin{equation}
\bar {\chi}(\beta) = \bar {\chi}_0 (1 - \beta/\beta_c)^{-\gamma}
            + \bar {\chi}_{\rm conf} (1 - \beta/\beta_c)^{-\gamma + \Delta_1}
            + \bar {\chi}_{\rm anal} (1 - \beta/\beta_c)^{-\gamma + 1}.
\label{eq:4.5}
\end{equation}
Including both correction terms
as free parameters into the fit routine is a hopeless enterprise. Instead,
we first tried fits with
$\beta_c$ held
{\em fixed} at values around 0.69290 in steps of 0.00001, and $\gamma,
\bar {\chi}_0,\bar {\chi}_{\rm conf}$, and $\bar {\chi}_{\rm anal}$ as
free parameters. The quality of the fits remained high
for a large range of $\beta$-values. The best fit with $Q = 0.97$ was
obtained for
$\beta_c = 0.69281$ and gave  $\gamma = 1.365(25)$, $\bar {\chi}_0 = 1.06(17)$,
$\bar {\chi}_{\rm conf} = -0.11(79)$, and
$\bar {\chi}_{\rm anal} = -0.25(1.18)$. We see
that the correction terms come out to be consistent with zero, but that the
error bars are fairly large due to the increased number of free parameters,
so that no definite conclusion can be drawn from these fits.

We therefore decided to
perform two further types of fits. First, we added to the leading term
in (\ref{eq:4.5}) only the analytic correction term (i.e.,
we enforced the amplitude of the confluent correction term to be zero),
and second, we tried to use only the confluent correction term
with fixed $\Delta_1=0.55$ (i.e., we put the amplitude of the analytic
correction equal to zero).
Fitting thus all 18 data points to the ansatz (\ref{eq:4.5})
with either $\bar {\chi}_{\rm conf}$
or $\bar {\chi}_{\rm anal}$ held fixed at zero we get for the other amplitude
$\bar {\chi}_{\rm anal} = -0.40(27)$ or $\bar {\chi}_{\rm conf} = -0.35(26)$,
respectively. The $Q$-value improves only marginally, and the amplitudes are
still almost consistent with zero. The other parameters change slightly, but
due to the much larger error bars than for the simple fit (\ref{eq:4.1}) they
are still compatible.

The latter fits are non-linear four-parameter fits
which are usually quite difficult to stabilize. To cope with this problem we
followed ref.~\cite{r9} and used the following method. For any pair of
values $\beta_c$, $\gamma$ we first minimized in the linear parameters
$\bar {\chi}_0$ and $\bar{\chi}_{\rm anal}$ (or $\bar{\chi}_{\rm conf}$),
which can be
done exactly (up to round-off errors). This yields a chi-squared function
$\chi^2(\beta_c,\gamma,\bar{\chi}_0(\beta_c,\gamma)$, $
\bar{\chi}_{\rm anal}(\beta_c,\gamma))$ $= \chi^2(\beta_c,\gamma)$
which depends on two non-linear parameters only
and can be minimized reliably by standard subroutines. Finally we used the so
determined estimates of $\beta_c,\gamma,\bar{\chi}_0,\bar{\chi}_{\rm anal}$
(or $\bar{\chi}_{\rm conf}$) as
initialization of the non-linear fit-routine MRQMIN of ref.~\cite{r24},
which yields then slightly improved parameter values and standard error
estimates. The fits with $\gamma, \bar{\chi}_0, \bar{\chi}_{\rm conf}$ and
$\bar{\chi}_{\rm anal}$ as free parameters were performed similarly.

One can also rewrite eq.~(\ref{eq:4.5}) as a function of $T$,
\begin{equation}
\bar {\chi}(T) = \bar {\chi}'_0 (T/T_c - 1)^{-\gamma}
        + \bar {\chi}'_{\rm conf} (T/T_c - 1)^{-\gamma + \Delta_1}
        + \bar {\chi}'_{\rm anal} (T/T_c - 1)^{-\gamma + 1},
\label{eq:4.6}
\end{equation}
with $\bar {\chi}'_0 = \bar {\chi}_0$, $\bar {\chi}'_{\rm conf} =
\bar {\chi}_{\rm conf}$, and
$\bar {\chi}'_{\rm anal} = \bar {\chi}_{\rm anal} + \gamma \bar {\chi}_0$.
This form is obtained from
(\ref{eq:4.5}) by a simple change of variable $\beta \to 1/T$, expanding
around $T_c$, and keeping the same correction terms. We believe that
(\ref{eq:4.6}) with $\bar {\chi}'_{\rm conf} = \bar {\chi}'_{\rm anal} = 0$
is, \'a priori, as justified as the simplest $\beta$-dependent ansatz
(\ref{eq:4.1}). Since HTS analyses are based on
expansions in $\beta$,  there is definitely a bias to prefer the ansatz
(\ref{eq:4.1}). We are not  aware, however, of a mathematically sound
justification for this choice. For what we think is a counter-example see
our correlation length analysis below.

Because the simplest $\beta$-dependent ansatz (\ref{eq:4.1}) works so good,
it is thus clear that for the temperature dependent fit ansatz we have
to take into account the analytic correction term.
As anticipated, trying to fit
the simplest ansatz (\ref{eq:4.6}) with
$\bar {\chi}'_{\rm conf} = \bar {\chi}'_{\rm anal} = 0$
to all 18 data points results in a fit with a comparatively
poor chi-squared of $\chi^2 = 21.66$, corresponding to $Q = 0.12$.
By again
successively discarding the data points with small correlation length, the
quality of the fit improves rapidly, and in Fig.~11 we see a weak trend that
the values of $\beta_c$ from the two types of fits may approach each other
asymptotically.

The fit to all 18 data points
with $\bar {\chi}'_{\rm anal}$ as a free parameter agrees perfectly with the
corresponding $\beta$-dependent ansatz (also with $\bar {\chi}_{\rm anal}$ as a
free parameter), yielding the expected correction
amplitude $\bar {\chi}'_{\rm anal} = 1.03(26)$. The fit with
$\bar {\chi}'_{\rm conf}$ as
free parameter is only slightly worse. The value of
$\bar {\chi}'_{\rm conf} = 0.90$
(instead of the expected result $\bar {\chi}'_{\rm conf} =
\bar {\chi}_{\rm conf} \approx 0$)
indicates, however, that in this case the confluent correction term trys
to account for the suppressed analytic correction term, and that it can
indeed mimic the analytic behavior very well.

The main result of all our efforts is that on the basis of standard statistical
tests, the simple ansatz (\ref {eq:4.1}) {\em turns out} to be justified.
We emphasize this point because \'a priori it is not at all clear that the
confluent correction is extremely small and that also the analytic correction
is negligible if the proper variable ($\beta$ in the case of $\bar {\chi}$)
is chosen. Although we feel that the corrections cannot
completely be ignored, one would need more accurate data
to account for these corrections in a reliable way.
\subsection{Correlation length}
We have repeated the same type of analysis for the correlation length data.
Starting again with the simplest possible ansatz taking into account only the
leading singularity, we find this time that the temperature dependent ansatz,
\begin{equation}
\xi(T) = \xi'_0 (T/T_c - 1)^{-\nu},
\label{eq:4.7}
\end{equation}
is well behaved. It should be emphasized that, as far as the role of $\beta$
and $T$ is concerned, this is just opposite to the situation for the
susceptibility. The fit to all 18 data points in Table~4 yields
\begin{eqnarray}
\beta_c  &=& 0.69281 \pm 0.00004, \label{eq:4.8a} \\
\nu      &=& 0.698   \pm 0.002,   \label{eq:4.8b} \\
\xi'_{0}  &=& 0.484   \pm 0.002    \label{eq:4.8c}
\end{eqnarray}
with a chi-squared of $\chi^2 = 8.14$, corresponding to a goodness-of-fit
parameter $Q = 0.92$, which may again be taken as \'a posteriori justification
of the simple ansatz (\ref{eq:4.7}). The estimate for
$\beta_c$ is somewhat smaller than the previous estimates, and also the
exponent $\nu$ is only barely compatible with our FSS value
$\nu = 0.704(6)$ and with the field theoretical estimates,
$\nu = 0.705(3)$ (resummed perturbation series \cite{zi80}),
$\nu = 0.710(7)$ (resummed $\epsilon$-expansion \cite{zi85}),
but it still lies within the $2\sigma$ error interval of these estimates.
If we extract $\xi$ from fits through the four lowest $\BF{k}$ values only,
we obtain the same central values as in (\ref{eq:4.8a})-(\ref{eq:4.8c}) with
about 1.5 times bigger error bars and a $Q$-value of 0.99. In Fig.~11 we
see that fits with the ansatz (\ref{eq:4.7}) are very stable against
discarding more and more data points with small correlation length.

Correspondingly, if we include into the $T$-dependent ansatz one correction
term at a time, as
described in the susceptibility analysis,
we obtain amplitudes that are fully consistent
with zero,  $\xi'_{\rm conf} = -0.048(65)$ and
$\xi'_{\rm anal} = -0.065(76)$.
For the corresponding
$\beta$-dependent fit we thus expect $\xi_{\rm conf} \approx 0$ and
$\xi_{\rm anal} = \xi'_{\rm anal} - \nu \xi'_0 = -0.267$. The latter
value is extremely well reproduced by the $\beta$-dependent fit with
analytic correction term ($\xi_{\rm anal} = -0.272(76)$).
Similar to the susceptibility the confluent
amplitude, however, comes out much too large in an attempt to mimic
the analytic correction term. We can thus conclude that also for the
correlation length confluent corrections are negligible. Analytic
correction terms are important in the $\beta$-dependent ansatz
but negligible in the $T$-dependent ansatz,
just opposite to the situation for the susceptibility.

It is noteworthy that, using
correction terms, the value of $\nu$ slightly increases, just opposite to
the case of the susceptibility data, where the value of $\gamma$ decreases
when corrections are included. It is impossible to find an objective criterion
for the systematic errors, but, with what we experienced, we feel that they are
at least of the order of the statistical errors, in particular for the
susceptibility, where the correction terms seem to be somewhat
more important than for the correlation length.

Combining the estimates for $\beta_c$ from the susceptibility and
correlation length
data in the high-temperature phase we obtain a final estimate of
\begin{equation}
\beta_c = 0.69288 \pm 0.00004,
\end{equation}
which is slightly smaller than our crossing value of $0.6930(1)$, but in
agreement with the MC estimate by Peczak {\em et al.} \cite{r18}.
\subsection{The exponent $\eta$}
Finally, we have combined the scaling behavior of $\xi$ and $\bar {\chi}$ to
get a direct estimate of the exponent $\eta = 2 - \gamma/\nu$ from
the relation $\bar {\chi} \propto \xi^{\gamma/\nu}$, or
\begin{equation}
\log \left(\bar {\chi}/\xi^2 \right) = c - \eta \log \xi,
\label{eq:4.8}
\end{equation}
where $c$ is a constant. By plotting $\bar {\chi}/\xi^2$ vs
$\xi$ on a log-log scale as in Fig.~12(a) we thus expect asymptotically
for large $\xi$
a straight line with slope $-\eta$. We see that up to $\xi \approx 12$
our data still give a curved line, indicating that corrections to asymptotic
scaling cannot yet be neglected. Taking a linear envelope to the last few
points as rough estimate, we obtain $\eta \approx 0.05$.
Alternatively we can define an effective exponent $\eta_{\rm eff}^{(2)}
= -\log \left( \bar {\chi}_{i+1} \xi_i^2 / \bar {\chi}_i \xi_{i+1}^2 \right)
/ \log \left( \xi_{i+1} / \xi_i \right)$ as the local slope between two
points in Fig.~12(a). Since this gives quite noisy results we have actually
computed the local slopes from linear fits through three and four points,
denoted by $\eta_{\rm eff}^{(3)}$ and $\eta_{\rm eff}^{(4)}$, respectively.
These effective exponents are plotted vs $\xi$ on a logarithmic scale
in Fig.~12(b). Again we see
that up to correlation lengths of the order of $\xi \approx 12$ the effective
exponents have clearly not yet reached the asymptotic value $( = \eta)$.
However, assuming a monotonic behavior as $\xi \rightarrow \infty$
we get at least an upper bound, $\eta < 0.05$.
Recall the field theory values which are
$\eta = 0.033(4)$ (resummed
perturbation series \cite{zi80}), $\eta = 0.040(3)$ (resummed
$\epsilon$-expansion \cite{zi85}\footnote{In a recent recalculation \cite{r99a}
of all Feynman graphs contributing to the highest order of the $\epsilon$
expansion ( ${\cal O}(\epsilon^5)$ ) several errors
were corrected. A subsequent reanalysis \cite{r99b} of the resummed series
gave a slightly smaller value for $\eta$. Compared with the error bar,
however, this change is negligible.}).

Estimates for the other critical exponents can be derived using (hyper-)
scaling
laws. We obtain $\beta = \frac {3} {2} \nu - \frac {1} {2} \gamma = 0.352(4)$,
and $\alpha = 2 - 3 \nu = -0.094(6)$ (compare also Table~5).
%
                       \section{Concluding remarks}
%
In conclusion we have shown that the {\em single\/}-cluster update
eliminates critical slowing down for the three-dimensional Heisenberg model
almost completely. As for the 3D Ising and XY model we expect that
it is more efficient than the {\em multiple\/}-cluster algorithm, but this has
not yet been explicitly verified due to the lack of data for the
multiple-cluster algorithm. Combined with
histogram reweighting and optimization techniques, finite-size scaling
analyses allow a precise
Monte Carlo determination of the critical exponents of the 3D Heisenberg model,
whose accuracy is comparable with the best estimates coming from field
theoretical methods. Direct analyses
of thermodynamic measurements
based on improved estimators in the high-temperature phase yield compatible
results. For the reader's convenience we have compiled some relevant
sources on the critical
parameters of the classical 3D Heisenberg model in Table~5.
Overall, our results are in good agreement with the
Monte Carlo values reported recently by Peczak {\em et al.} \cite{r18}.
In particular we confirm that the value for
$\beta_c$ on a simple cubic lattice is significantly higher than previous
estimates coming from analyses of
high-temperature series expansions. We would like to remark, however, that a
very recent analysis \cite{ahj92} of extended series expansions \cite{lw}
(up to 14th order), using more refined Pad\'e-approximant techniques,
is consistent with the
MC estimates.
%
                 \section*{Acknowledgement}
%
The numerical simulations were performed on the CRAY X-MP and Y-MP of
the Konrad-Zuse Zentrum f\"ur Informationstechnik Berlin (ZIB),
the CRAY X-MP at the Rechenzentrum der Universit\"at Kiel and, in the late
stage of the project, also on the CRAY Y-MP of the
H\"ochstleistungsrechenzentrum (HLRZ), Forschungszentrum J\"ulich.
We thank all institutions for their generous support.
%
     
%
%
\newpage
%
%
%
 \begin{table}              
{\Large\bf Tables}\\[1cm]
 \begin{center}
  \begin{tabular}{|r|c|c|c|c|c|}
   \hline
\multicolumn{1}{|r|} {$L$} &
\multicolumn{1}{c|}{$\beta$} &
\multicolumn{1}{c|}{$N$}  &
\multicolumn{1}{c|}{$f$}  &
\multicolumn{1}{c|}{$\langle |C| \rangle$}  &
\multicolumn{1}{c|}{$\tau$}  \\ \hline
%
 12 & 0.6783 & 1430249 & 0.045 & 79 & 0.8 \\
 12 & 0.6929 & 5222274 & 0.071 & 122 & 1.2 \\
 12 & 0.7009 &  931277 & 0.086 & 149 & 1.4 \\
\hline
 16 & 0.6872 &  242804 & 0.122 & 167 & 1.4 \\
 16 & 0.6929 &  217393 & 0.157 & 213 & 1.8 \\
 16 & 0.6953 &  206676 & 0.174 & 241 & 1.9 \\
\hline
 20 & 0.6862 &  270067 & 0.110 & 220 & 1.1 \\
 20 & 0.6929 &  332725 & 0.208 & 332 & 1.8 \\
 20 & 0.6947 &  284544 & 0.184 & 369 & 1.8 \\
\hline
 24 & 0.6872 &  709711 & 0.130 & 300 & 1.0 \\
 24 & 0.6929 &  472831 & 0.207 & 476 & 1.6 \\
 24 & 0.6953 &  350379 & 0.244 & 563 & 1.7 \\
 24 & 0.6972 &  355572 & 0.274 & 631 & 1.7 \\
\hline
 32 & 0.6881 &  356881 & 0.126 & 460 & 1.1 \\
 32 & 0.6929 &  351394 & 0.206 & 842 & 1.8 \\
 32 & 0.6965 &  360726 & 0.215 & 1174 & 1.9 \\
\hline
 40 & 0.6904 &  255563 & 0.106 & 851 & 1.1 \\
 40 & 0.6929 &  216251 & 0.163 & 1304 & 1.5 \\
 40 & 0.6953 &  232234 & 0.194 & 1774 & 1.8 \\
\hline
 48 & 0.6914 &  131620 & 0.145 & 1339 & 1.2 \\
 48 & 0.6929 &  138729 & 0.200 & 1840 & 1.6 \\
 48 & 0.6941 &  126643 & 0.165 & 2287 & 1.6 \\
 48 & 0.6968 &  174441 & 0.178 & 3287 & 1.7 \\
\hline
\end{tabular}
\end{center}
\caption[a]{Measurement statistics: $N$ is the number of measurements,
 taken after
$f \times L^3$ spins are flipped (on the average), $\langle |C|
\rangle$ is the average cluster size that
is measured after each update step, and $\tau$ is a rough estimate of the
autocorrelation time in Metropolis-equivalent units.}  \end{table}
\clearpage
\newpage
%
\vspace{0.50in}
 \begin{table}[t]              
 \begin{center}
  \begin{tabular}{|l|c|r|r|l|l|r|r|}
   \hline
 & \multicolumn{2}{c|}{$\chi$} &
\multicolumn{2}{c|}{$\chi^{\rm c}$} & &
\multicolumn{2}{c|}{$\chi^{\rm c}_{\rm max}$} \\ \hline
\multicolumn{1}{|c|}{$\beta$} & $\eta$ &
\multicolumn{1}{c|}{$Q$}  &
\multicolumn{1}{c|}{$\eta$}  &
\multicolumn{1}{c|}{$Q$}  & &
\multicolumn{1}{c|}{$\eta$}  &
\multicolumn{1}{c|}{$Q$}
  \\ \hline
0.6929  & 0.0364(17) & $0.36$ &  0.0086(44) & 0.71 & & 0.0231(61) &  0.30
  \\
0.6930  & 0.0271(17) &  $0.78$ &  0.0156(44) & 0.69 &  & &
  \\
0.6931  & 0.0178(17) &  $0.43$ &  0.0237(44) & 0.48 &  & &
  \\
   \hline
  \end{tabular}
 \end{center}
 \caption[a]{Results for $\eta$ obtained from
log-log fits
of $\chi / L^2$ and
$\chi^{\rm c}/ L^2$ versus the lattice size $L$ at
three temperatures near $\beta_c$. Included is also the estimate of $\eta$
coming
from a fit of the
maximum $\chi^{\rm c}_{\rm max}$ of the connected susceptibility (see text).
$Q$
denotes the quality factor of the least-square fit routine FIT of
ref.~\cite{r24}.}
\end{table}
%

\begin{table}[b]             
 \begin{center}
  \begin{tabular}{|r|c|c|c|c|}
   \hline
\multicolumn{1}{|r|} {$L$} &
\multicolumn{1}{c|}{$T_{\chi_{\rm max}^{\rm c}}$} &
\multicolumn{1}{c|}{$\chi_{\rm max}^{\rm c}$}  &
\multicolumn{1}{c|}{$T_{C_{\rm max}}$}  &
\multicolumn{1}{c|}{$C_{\rm max}$} \\ \hline
%
 12 & 1.4747(35) & 5.260(38)  & 1.4243(101) & 2.426(11) \\
 16 & 1.4629(28) & 9.242(100) & 1.4362(45) & 2.579(26)  \\
 20 & 1.4579(13) & 14.25(10)  & 1.4393(39) & 2.711(20)  \\
 24 & 1.4550(11) & 20.80(09)  & 1.4355(37) & 2.778(15)  \\
 32 & 1.4505(11) & 36.53(22)  & 1.4386(35) & 2.936(34) \\
 40 & 1.4486(08) & 56.45(39)  & 1.4430(21) & 2.959(46) \\
 48 & 1.4474(06) & 81.48(63)  & 1.4415(15) & 3.090(50) \\
\hline
\end{tabular}
\end{center}
\caption[a]{Data for the temperature locations and the absolute values of the
maxima of the magnetic susceptibility and the specific heat. The data was
extracted from the optimally combined curve of the three runs for each lattice
size (compare text).}
\end{table}
\clearpage
\newpage
%
%
%
%
\addtolength{\oddsidemargin}{-2.6cm}
\setlength{\evensidemargin}{\oddsidemargin}
 \begin{table}[h]              
 \begin{center}
  \begin{tabular}{|l|r|c|r|c|c|c|r|r|r|}
   \hline
\multicolumn{1}{|c|}{$\beta$} & $L$ &
\multicolumn{1}{c|}{$L/\xi_{\rm imp}$}  &
\multicolumn{1}{c|}{$N/10^3$}  &
\multicolumn{1}{c|}{$f$}  &
\multicolumn{1}{c|}{$\langle |C| \rangle/\bar {\chi}_{\rm imp}$}  &
\multicolumn{1}{c|}{$C$}  &
\multicolumn{1}{c|}{$\bar {\chi}$}  &
\multicolumn{1}{c|}{$\bar {\chi}_{\rm imp}$}  &
\multicolumn{1}{c|}{$\xi_{\rm imp}$}  \\ \hline
0.650 & 32 & 9.9 & 79.5 & 0.270 & 0.7556 & 1.074(13) &~46.00(15) &~45.697(50)
&~3.2345(26) \\
0.655 & 32 & 9.0 &263.6 & 0.134 & 0.7547 & 1.137(11) &~54.43(12) &~54.280(53)
&~3.5412(24) \\
0.660 & 32 & 8.1 & 93.0 & 0.271 & 0.7539 & 1.222(13) &~65.95(21) &~66.095(89)
&~3.9336(34) \\
0.665 & 40 & 9.0 &400.0 & 0.133 & 0.7531 & 1.310(11) &~83.07(14) &~83.155(65)
&~4.4343(24) \\
0.670 & 40 & 7.8 &138.8 & 0.134 & 0.7523 & 1.382(19) &109.32(33) &109.59(18)
&~5.1260(53) \\
0.673 & 50 & 8.8 & 89.5 & 0.133 & 0.7519 & 1.453(23) &132.77(47) &132.76(22)
&~5.6676(62) \\
0.675 & 50 & 8.2 &114.4 & 0.132 & 0.7517 & 1.470(24) &153.92(48) &153.99(26)
&~6.1197(64) \\
0.676 & 60 & 9.4 &132.3 & 0.133 & 0.7515 & 1.539(21) &167.07(47) &166.69(21)
&~6.3740(55) \\
0.677 & 60 & 9.0 & 68.5 & 0.133 & 0.7515 & 1.560(29) &181.71(72) &181.16(34)
&~6.6584(83) \\
0.678 & 60 & 8.6 &170.5 & 0.133 & 0.7513 & 1.562(20) &198.52(49) &198.74(25)
&~6.9869(58) \\
0.679 & 60 & 8.2 &156.3 & 0.133 & 0.7512 & 1.619(21) &218.82(58) &218.46(32)
&~7.3333(68) \\
0.680 & 60 & 7.8 & 95.3 & 0.162 & 0.7511 & 1.700(28) &242.04(79) &242.17(44)
&~7.7359(87) \\
0.681 & 70 & 8.5 &295.5 & 0.133 & 0.7510 & 1.710(17) &270.68(51) &270.96(27)
&~8.2013(52) \\
0.682 & 70 & 8.0 &154.9 & 0.133 & 0.7508 & 1.776(23) &304.38(79) &305.91(46)
&~8.7244(81) \\
0.683 & 80 & 8.6 & 73.4 & 0.157 & 0.7508 & 1.765(31) &348.5(1.3) &349.26(63)
&~9.346(11)~~ \\
0.684 & 80 & 7.9 &138.4 & 0.133 & 0.7506 & 1.778(25) &405.9(1.2) &405.92(62)
&10.0988(94)  \\
0.685 & 90 & 8.2 &174.0 & 0.133 & 0.7505 & 1.903(24) &477.9(1.2) &478.60(64)
&10.9857(92)  \\
0.686 &100 & 8.3 & 77.4 & 0.133 & 0.7504 & 1.860(36) &575.0(2.0)
&$\!576.9(1.1)$  &$\!12.093(15)$~  \\
\hline
\end{tabular}
\end{center}
\caption[a]{Results in the high-temperature phase. $N$ is the number of
measurements of the non-improved observables, taken after $f \times L^3$ spins
are flipped (on the average). $C$ denotes the specific heat and
$\langle |C| \rangle$ is the average cluster size. The cluster size
and the improved observables are measured after each cluster-update step.}
\end{table}
\clearpage
\newpage
\addtolength{\oddsidemargin}{2.6cm}
\setlength{\evensidemargin}{\oddsidemargin}
%
%
%
\addtolength{\oddsidemargin}{-1.0cm}
\setlength{\evensidemargin}{\oddsidemargin}
\begin{table}[t]             
\newlength{\digitwidth} \settowidth{\digitwidth}{\rm 0}
\catcode`?=\active \def?{\kern\digitwidth}
 \begin{center}
  \begin{tabular}{|c|c|c|c|c|c|}
   \hline
\multicolumn{1}{|c|} {}  &
\multicolumn{2}{c|} {field theory} &
\multicolumn{3}{c|} {MC simulations} \\
\multicolumn{1}{|c|} {critical}  &
\multicolumn{1}{c|} {$g$-expansion } &
\multicolumn{1}{c|} {$\epsilon$-expansion } &
\multicolumn{1}{c|}{Metropolis MC }  &
\multicolumn{2}{c|}{this study } \\
\multicolumn{1}{|c|} {parameter} &
\multicolumn{1}{c|} {ref.~\cite{zi80}} &
\multicolumn{1}{c|} {refs.~\cite{br85,zi85}} &
\multicolumn{1}{c|}{ref.~\cite{r18}}  &
\multicolumn{1}{c|}{ FSS}  &
\multicolumn{1}{c|}{ HT} \\
\hline
%
$\beta_c$      & -               & -              & ?0.6929(1)     &
  ?0.6930(1)     & ?0.69288(4) \\
$U^*$          & -               & $\!0.59684$      &  ?0.622(1)?    &
  ?0.6217(8)     & - \\
$\nu$          & ?0.705(3)??     & ?0.710(7)?     & ?0.706(9)?     &
  ?0.704(6)?     & ?0.698(2)?? \\
$\alpha$       & $\!-0.115(9)??$ & $\!-0.130(21)$ & $\!-0.118(18)$ &
  $\!-0.112(18)$ & $\!-0.094(6)??$ \\
$\alpha / \nu$ & $\!-0.163(12)?$ & $\!-0.183(28)$ & $\!-0.167(36)$ &
  $\!-0.159(24)$ & $\!-0.135(9)??$ \\
$\beta$        & ?0.3645(25)     & ?0.368(4)?     & ?0.364(7)?     &
  ?0.362(4)?     & ?0.352(4)?? \\
$\beta / \nu$  & ?0.517(6)??     & ?0.518(11)     & ?0.516(3)?     &
  ?0.514(1)?     & ?0.504(5)?? \\
$\gamma$       & ?1.386(4)??     & ?1.390(10)     & ?1.390(23)     &
  ?1.389(14)     & ?1.391(3)?? \\
$\eta$         & ?0.033(4)??     &  ?0.040(3)?    & ?0.031(7)?     &
  ?0.027(2)?     & $< 0.05$ \\
\hline
\end{tabular}
\end{center}
\caption[a]{Various sources for estimates of the critical
parameters for the classical 3D Heisenberg model. For the MC simulations
scaling
relations were used to obtain estimates for $\alpha$ (all) and $\beta$
(only HT), and the FSS values of the exponents $\gamma$ and
$\beta$  were calculated from the measured ratios, using the estimate for
$\nu$.
The field theory estimates of ratios with $\nu$ are calculated from the
values and errors of the critical exponents.}
\end{table}  
\clearpage
\newpage
\addtolength{\oddsidemargin}{1.0cm}
\setlength{\evensidemargin}{\oddsidemargin}
%
  {\Large\bf Figure Headings}
  \vspace{1in}
  \begin{description}
    \item[\tt\bf Fig. 1:]
(a) The energy histograms for $L=48$ at the three simulation temperatures.
(b) The constant energy averages $\langle\!\langle m
\rangle\!\rangle (E)$ as computed from the three simulations yielding the
histograms in (a).
(c) The weights for the optimal combination of the histograms according to
the procedure in ref.~\cite{r21}.
    \item[\tt\bf Fig. 2:]
The Binder parameter $U_L$ vs $\beta$. The values of $U_L(\beta)$ were
obtained by reweighting and optimized combining the results of our
three simulations at different
temperatures for each lattice size $L$.
The simulations were performed at
$\beta_0=0.6929$ (the critical inverse temperature found by
Peczak {\em et al.\/} \cite{r18}),
and at the positions of the maxima of the specific heat $C$ and
the susceptibility $\chi^{\rm c}$, respectively.
    \item[\tt\bf Fig. 3:]
Estimates for $T_c$, coming from plotting the $T$-values of the crossing
points of the $L$=12 and $L$=16
curves of the Binder parameter $U_L$ vs the inverse logarithm of the
 scale factor
$b = L'/L$. The extrapolation leads to an estimate of
$\beta_c = 0.6930(1)$.
    \item[\tt\bf Fig. 4:]
The thermodynamic  derivative $\frac {dU_L} {d\beta}$ calculated at
$\beta_c$=0.6930 vs lattice size $L$
in a double logarithmic plot. The slope of the linear
least-square fit gives an estimate for the
critical exponent of the correlation length, $\nu = 0.704(6)$, with a
quality $Q$=0.61.
    \item[\tt\bf Fig. 5:]
Double logarithmic plot of the magnetization $\langle m \rangle$ at
$\beta_c = 0.6930$ vs lattice size $L$. The
values of $\langle m \rangle$ were
obtained by reweighting and optimized combining (see text) of our three
simulations at different
temperatures for each lattice size $L$. The slope gives an estimate for
$\beta/\nu$. From the fit we
obtain $\beta/\nu=0.514(1)$, with quality $Q=0.68$.
    \item[\tt\bf Fig. 6:]
log($\chi/L^2)$ vs log($L$) at inverse temperatures
$\beta$=0.6925 ($\Box$),
$\beta$=0.6927 ($\Diamond$),
$\beta$=0.6929 ($\Diamond$),
$\beta$=0.6930 ($\times$),
$\beta$=0.6931 ($\bigtriangleup$),
$\beta$=0.6933 ($\bigtriangleup$), and
$\beta$=0.6935 ($\circ$).
The straight line corresponds to a linear least-square fit at
$\beta$=0.6930, that gives  $\eta$=0.0271(17) with $Q$=0.78.
    \item[\tt\bf Fig. 7:]
Variation of the pseudo transition temperatures $T_{\chi^{\rm c}_{\rm max}}(L)$
and $T_{\rm C_{\rm max}}(L)$  with $L^{-1/\nu}$, where $\nu=0.704(6)$ is our
estimate obtained in
Fig.~4. The fits yield estimates of $\beta_c = 0.6930(3)$ ($Q=1.0$) and
 $\beta_c = 0.6925(9)$ ($Q=0.80$), respectively.
    \item[\tt\bf Fig. 8:]
The $Q$-value (quality-factor) of linear least-square log-log fits of the
quantities $\chi (\beta)$, $\chi^{\rm c} (\beta)$, $\langle m \rangle (\beta)$,
and $\frac {dU_L} {d\beta} (\beta)$ vs the lattice size $L$,
extracted at different $\beta$-values in intervals of $\Delta\beta$=0.00002.
    \item[\tt\bf Fig. 9:]
Lattice-size dependence of the maxima of the specific heat, $C_{\rm max}$.
The solid curve
is a non-linear three-para\-me\-ter fit, whereas the dashed curve comes from a
linear fit
with fixed $\alpha/\nu = 2/\nu - D = -0.159$, employing hyperscaling arguments
and our estimate of $\nu = 0.704(6)$.
%
    \item[\tt\bf Fig. 10:]
Fit of $\hat{G}(\BF{k})^{-1}$ to $c [ \sum_{i=1}^3 2(1-\cos k_i) + (1/\xi)^2]$
on a $100^3$ lattice used to compute the correlation length $\xi$.
    \item[\tt\bf Fig. 11:]
The critical coupling $\beta_c$ as estimated from non-linear
three-para\-me\-ter
fits to $\xi$ and $\bar {\chi}$ assuming the leading power-law singularity
only,
written as function of $\beta$ or $T$. The $x$-axis shows how many data point
are taken into account in these fits (i.e., from right to left more and more
points with small correlation length are discarded). In order to disentangle
the error bars two curves are slightly displaced in $x$.
    \item[\tt\bf Fig. 12:]
(a) Log-log plot of $\bar {\chi}/\xi^2$ vs $\xi$. The slope of the linear
envelope for large $\xi$ gives $\eta \approx 0.05$.
(b) The effective critical exponent $\eta_{\rm eff}$ vs $\xi$ on a logarithmic
scale, using two different discretization schemes. The asymptotic value of
$\eta_{\rm eff}$ as $\xi \rightarrow \infty$ is an estimate for the
critical exponent $\eta = 2 - \gamma/\nu$.
  \end{description}
\end{document}